\documentclass[aps,12pt,showkeys,amsmath,amsfonts,amssymb,
notitlepage]{revtex4-1}

\usepackage[dvips]{graphicx}

\usepackage[english]{babel}
\usepackage{epstopdf}
\DeclareMathOperator{\arcsinh}{arcsinh}

\begin{document}

	\title{Centrifugal cosmological repulsive force\\in a homogeneous universe}
	\author{\firstname{A.~V.}~\surname{Klimenko}}
	\email{alklimenko@gmail.com}
	\affiliation{``Business and Technology'', Chelyabinsk, Russia}
	\author{V.~A.~Klimenko}
	\affiliation{``Chelyabinsk state university'', Chelyabinsk, Russia}
		
\begin{abstract}
We study the dynamics of homogeneous isotropic three-dimensional worlds filled with radiation ($3\mathbb R$-worlds).
It is shown that the dynamics of these worlds with the additional fourth large-scale spatial dimension leads to an important effect.
At $3\mathbb R$-worlds the forces of repulsion appear.
The source of these forces is the thermal energy of the radiation that fills these worlds.
In the four-dimensional space, these forces are centrifugal.
They operate in an external for $3\mathbb R$-world spatial dimension and stretch it.
In the three-dimensional comoving coordinate system the centrifugal forces shows themselves as forces of repulsion.
Standard Einstein's equations do not describe these forces.

Written generalized Einstein's equation describing the dynamics of a homogeneous isotropic universe, taking into consideration the centrifugal forces of repulsion.
We propose a cosmological model of the universe, based on these equations.
This model apply to explain the observation data.
	

\end{abstract}	

\maketitle
	
This work is a refinement and development of research on the nature of the cosmological repulsion forces carried out earlier us with Academician of Russian Academy of Sciences A.M.~Friedman.
This work is dedicated to the memory of this remarkable man and famous Russian physicist.


\section{Introduction}
\label{vvedenie}
Exist yet recently believed that the dynamics of the universe is determined by the gravitational force (see, eg, \cite{1,2,19}).
Currently, there are numerous observational data to explain which is necessary to assume the existence of a cosmological repulsion forces.
There is good reason to believe that the dynamics of the universe is largely determined by the influence of these forces (see, eg, \cite{3,4,5,6,7}).
	
Prevailing opinion that the cosmological repulsion force is what is described by $\Lambda$-term in Einstein's equations (see, eg, \cite{8,9}).

Introduction $Lambda$-term in the equations of general relativity (GR) in \cite{10} was connected with Einstein's ideas about the properties of the universe.
In writing this work he was convinced that the universe is not only homogeneous and isotropic, but also stationary. 
To explain the stationarity of the universe Einstein had assumed that, along with the forces of gravity in nature, there is also a repulsive force.
He believed that the equilibrium of these forces provides the stationarity of the universe.
connect cosmological repulsion forces with known properties of matter, Einstein could not.
He suggested that these forces are related to something else.
It's something ($\Lambda$-term) creates unremovable curvature of space-time, which is the source of the cosmological repulsion forces. 
	
Later, after the analysis of observations, it became clear that the universe is not stationary, Einstein had lost interest in the problem of the cosmological repulsion forces.
Why introduce into GR $\Lambda$-term, if nonstationarity of the universe can be explained without it?
	
The situation changed dramatically in the last decade.

It became clear that to explain many of the observational data is necessary to assume the existence of the cosmological repulsion forces. 
The popularity of the concept of $\Lambda$-term again rose sharply.
It is often argued that the $\Lambda$-term yields a correct description of the cosmological repulsion forces (see, for example \cite{8,9}).
	
Common is the assertion that the source of the cosmological forces of repulsion, described by $\Lambda$-term (hereinafter called Einsteinian), is a vacuumlike medium called ``dark energy'' \cite{8,9,11,12}.
Believe that the density of energy  $\varepsilon_\Lambda$ and the pressure $P_\Lambda$ of ``dark energy''
is defined by:
\begin{equation}
	\label{P_Lambda}
		\varepsilon_\Lambda = c^4 \Lambda / 8 \pi G,\;\;P_{\Lambda}=-\varepsilon_{\Lambda},
	\end{equation}
where $G$~--~the gravitational constant, $c$~--~the speed of light, $\Lambda$~--~cosmological constant.
It is assumed that the $\Lambda>0$.

We can show that the formula defining the cosmological acceleration produced by ``dark energy'' in a homogeneous isotropic universe is (see, e.g. \cite[ch.4]{1}):
\begin{equation}
	\label{d2a}
		\ddot{a}_\Lambda =- \frac {4}{3} \pi G \frac {a}{c^2} (\varepsilon_\Lambda + 3 P_\Lambda) =\frac {1}{3} \Lambda c^2 a,
	\end{equation}
where $a(t)$~--~radius of curvature of the universe.
Dot denotes time derivative.

For Einsteinian repulsive force, whose source is $\varepsilon_\Lambda$ and $P_\Lambda$, it is important that $\varepsilon_\Lambda>0$ and $P_\Lambda=-\varepsilon_\Lambda<0 $, therefore these forces arise.
Also, we should pay attention to the fact that field of Einsteinian forces exist even if $G\equiv 0$.
Consequently, naturally consider that this field is independent of the gravitational field.
It is also natural to assume that the source of Einsteinian repulsive force is not a hypothetical ``dark energy'', whose parameters depend on the gravitational constant $G$ (see formula \eqref{P_Lambda}), but something else, does not depend on $G$ .
	
In the papers \cite{13, 14} are considered models of homogeneous centrosymmetric infinite gravitating systems.
They are a variant of the cosmological models (``World on the brane''), in which the universe is regarded as a three-dimensional brane in four-dimensional space \cite{15}.
However, in contrast to previous studies (see, eg, \cite {16}), dedicated to the development of various aspects of this model in the works \cite{13, 14} into consideration not only normal to the brane velocity of the space environment, but also the tangential velocity .
Consideration of the tangential velocity allows us to describe the centrifugal forces acting on each element of the brane in the direction of the external spatial dimension to it.
Important in these studies is the assumption of the existence of such a measurement.
From the perspective of a typical observer on the brane, these forces appear as the cosmological repulsion forces.
In this paper we continue the study of the dynamics of the idealized homogeneous isotropic worlds filled with radiation, which was begun in \cite{13, 14}.
	
Given a clear explanation of the kinematic cosmological repulsive forces in these worlds.
It is shown that they are centrifugal in nature.
Written generalized Einstein's equations, taking into consideration these forces.
The source of these forces is the thermal energy of the space matter.
In the universe the vast majority of this energy is concentrated in a uniformly distributed in the space of the relativistic component of the space matter.
The equations describing dynamics of the homogeneous isotropic universe taking into consideration centrifugal forces are obtained.
To show that the proposed explanation of the cosmological repulsion forces has to do with reality, the equations used to explain the observational data.
	
The work is structured as follows.
In section \ref{dinamika_onor_rel_mirov}, considered the dynamics of the relativistic homogeneous isotropic worlds.
It is shown that the repulsive forces in these worlds are centrifugal in nature and are not described in the standard Einstein's equations.
Generalized Einstein equation, allowing to describe these forces in homogeneous isotropic environments are recorded in section \ref{obob_urav_einst}.
This section also is given generalized cosmological equations Friedman's, describing the dynamics of the universe, taking into consideration the action of centrifugal forces.
In section \ref{rel_model} proposed kinematic cosmological model of the universe.
Examples of using the proposed cosmological model to explain some important cosmological observational data is contained in section \ref{prilogenie}.
In section \ref{zakl} is a short list of the results.
Appendix contains a study of the dynamics of an idealized non-relativistic universe, taking into account the action of centrifugal forces.

\section{The dynamics of homogeneous isotropic relativistic worlds}
\label{dinamika_onor_rel_mirov}

\subsection{Problem Statement}
We study the dynamics of an idealized isotropic homogeneous three-dimensional world filled with radiation.
further, this world is called by $3\mathbb R$-World ($\mathbb R$~--~Relativistic).
Initially, for simplicity, research was conducted under the assumption that the action of cosmological gravitational field be absent.
For clarity, $3\mathbb R$-world view as a three-dimensional homogeneous isotropic nonstationary hypersphere radius of curvature of which $a(t)$ is the same at all points.
The final description of the dynamics of $3\mathbb R$-worlds considered into account the influence of the gravitational field, as well as take into account the existence possibility of $3\mathbb R$-worlds with negative curvature (see \ref{rel_model}).
We believe that a spherical $3\mathbb R$-world is inside in a  four-dimensional flat Euclidean space.
To study the dynamics of $3\mathbb R$-world use two coordinate systems: four-dimensional spherical $K_4{(r,\chi,\theta,\varphi)}$, with origin at the center of the hypersurface, as well as the accompanying three-dimensional time-dependent polar coordinate system $K_3{(R_3=a_3(t_3),\chi,\theta,\varphi)}$.
Length $a_3(t_3)$ is the scale factor of $K_3$.
Here and below the icons ``3'' and ``4'' denote quantities relating to systems of coordinates $K_3$ and $K_4$, respectively.
Consider two cases.
In the first case $3\mathbb R$-world is cold, and in the second case $3\mathbb R$-world is hot.

\subsection{Dynamics of cold $3\mathbb R$-world}
Assume that the initial time $t=0$ in the system $K_4$ particles of radiation, further, for brevity, called photons, are on a three-dimensional surface of hypersphere  of radius $a_m$, uniformly fill it and move strictly radially from its center.
Since all the photons at any point of $K_4$ move identically, then the radiation that fills $3\mathbb R$-world is cold.

Obviously, for $t>0$ in the coordinate system $K_4$ $3\mathbb R$-world uniformly expanding with the speed of light $c$.
Equation describing its dynamics has the form:
\begin{equation}\label{1}\frac{d^2a_4}{dt_4^2}=0.\end{equation}

Solution of equation \eqref{1}, satisfying the initial conditions
\begin{equation}\label{2}a_4(0)=a_m,\;\frac{da_4(0)}{dt_4}=c,\end{equation}
is a function:
\begin{equation}\label{3}a_4(t_4)=a_m+c\,t_4.\end{equation}

It is obvious that the evolution of cold $3\mathbb R$-world begins at $t_4=-a_m/c$ from the singular state.
Using the accompanying coordinate system $K_3$, ``frozen'' in the expanding cool $3\mathbb R$-world is impossible.
This case for introduction of attendant coordinate system is degenerate.

\subsection{Dynamics of Hot $3\mathbb R$-world}
Assume that in the initial time $t_0=0$ $3\mathbb R$-world was motionless and had a minimum size of $a_m$.
This means that in a coordinate system $K_4$ all the photons initially located on a hypersphere of radius $a_m$ and had only a tangential velocity (${V_r(0)=0},\;{V_t(0)=c}$).

The initial state $a(0)=a_m$ of $3\mathbb R$-world is not in equilibrium.
Consequently, the $ 3\mathbb R $ - world is expanding and expanding is uneven.
Consider its dynamics in the systems of coordinates $K_4$ and $K_3$.

\subsubsection{Dynamics in a coordinate system $K_4$}
\label{subsub_1}
The function $a_4 (t_4)$, describing the dynamics of the hot $3\mathbb R$-world in coordinate system $K_4$ defined by:
\begin{equation}\label{4}a_4(t_4)=\sqrt{a_m^2+c^2t_4^2}.\end{equation}

It is a solution of equation
\begin{equation}\label{5}\frac{d^2a_4}{dt_4^2}=\frac{a_m^2c^2}{a_4^3}\end{equation}
with initial conditions
\begin{equation}\label{6}a_4(0)=a_m,\;\;\frac{da_4(0)}{dt_4}=0.\end{equation}

In the system of $K_4$ tangential component of the velocity of photons at a distance $a_4(t_4)$, defined by equation:
\begin{equation}\label{7}V_t(t_4)=c\,a_m/a_4(t_4).\end{equation}

Taking into account \eqref{7}, the equation \eqref{5} written in the form:
\begin{equation}
	\label{8}
	\frac{d^2a_4}{dt_4^2}=-\frac d{da_4}\left(\frac{V_t^2}2\right)=\frac{V_t^2}{a_4}.
\end{equation}

Formally \eqref{8} can be regarded as an equation describing in the framework of mechanics of continuum medium expansion of $3\mathbb R$-world under the forces of repulsion.
As seen from \eqref{8}, the source of these forces in the system $K_4$ is the energy
\begin{equation}
	\label{9}
	V_t^2/2=\frac{c^2a_m^2}{2\,a_4^2}.
\end{equation}

Repulsive forces are centrifugal in its nature.
The origin of these forces ``purely geometric''.
In the problem of motion in the central field (see, eg, \cite[\S14]{17}) the energy, defined by \eqref{9}, is called centrifugal.
In the problem considered here, this term should be retained.

Taking into account isotropic distribution of tangential velocities, the energy $V_t^2/2$ can also be considered as the thermal energy, and the equation \eqref{9} as the first law of thermodynamics.
It is possible to show that it describes the adiabatic expansion of a relativistic gas with adiabatic exponent $\gamma=4/3$.
In a coordinate system $K_4$ this gas is cold in the radial direction and hot in the tangential.

As part of the standard cosmological equations of Einstein centrifugal repulsive force in the $3\mathbb R$-world can not be described (see para \ref{st_ur_fridman}).
To describe them, the standard Einstein's equations must be generalized.
The adjusted technique of introducing cosmological repulsion forces into the equations of Einstein's, previously described in \cite{20}, for the convenience of readers, placed in section \ref{obob_urav_einst} of present work.

Five-dimensional space-time associated with the coordinate system $K_4$, is flat.
Geodesic in this space are straight lines.
Each particle is $3\mathbb R$-world in the $K_4$ moves uniformly and rectilinearly.
At the same time, the four-dimensional space-time associated with the coordinate system $K_3$ is a curved and expands unevenly.

\subsubsection{The dynamics in the reference system $K_3$}
To describe the dynamics of the $3\mathbb R$-world in the system $K_3$ convenient to use concept ``typical observers''.
Typical observers --- it is abstract identical objects which uniformly fills $3\mathbb R$-world and moving radially in the system $K_4$ at speeds $da_4/dt_4$.
We assume that with any of the typical observer can be attributed to his proper one-dimensional rectilinear coordinate system $K_{34}$, moving along the radial coordinate system $K_4$.
A typical observer is at the origin of $K_{34}$ and moves in the system $K_4$ at speeds $da_4/dt_4$.
We consider that the proper time $t_{34}$ of a typical observer in the system $K_{34}$ and time $t_3$, the accompanying system of coordinates $K_3$, are identical concepts, and more times $t_{34}$ and $t_3$ do not differentiate.

Taking into account the invariance of the interval between two infinitesimally close events we find the relation between time $t_4$ of the system $K_4$ and time $t_3$ of the system $K_3$.
Let $dt_3$ is the time between events occurring with a typical observer in a system $K_3$, and $dt_4$ in a system $K_4$.
Equality:
\begin{equation}
	\label{10}
	c^2dt_3^2=c^2dt_4^2-da_4^2,
\end{equation}
where $da_4$~--~ distance at which a typical observer displaced from the center of $3\mathbb R$-world in a system $K_4$ during $dt_4$ between these events.
From \eqref{10} we find:
\begin{equation}
	\label{11}
	dt_3=dt_4\sqrt{1-\frac{\left(da_4/dt_4\right)^2}{c^2}}.
\end{equation}

Considering \eqref{4}, receive:
\begin{equation}
	\label{17}
	\frac{da_4}{dt_4}=\frac{c^2t_4}{\sqrt{a_m^2+c^2t_4^2}}.
\end{equation}

From \eqref{11}, taking into account \eqref{17}, we find:
\begin{equation}
	\label{18}
	dt_3=\frac{a_mdt_4}{\sqrt{a_m^2+c^2t_4^2}}.
\end{equation}

Integrating \eqref{18}, and assuming that the time ${t_4=0}$ corresponds to the time ${t_3=0}$, we find:
\begin{equation}
	\label{19}
	t_3=t_m\arcsinh{\frac{t_4}{t_m}},
\end{equation}
where ${t_m=a_m/c}$.
This formula defines the correspondence between indications $t_3$ of clocks of a typical observer, and a indications $t_4$ of clocks of system $K_4$, by which it flies.

Given that ${da_4/dt_4=\left(da_4/dt_3\right)\cdot\left(dt_3/dt_4\right)}$ and using \eqref{11} we find:
\begin{equation}
	\label{555}
	\frac{da_4}{dt_3}=\frac{da_4}{dt_4}\left/\sqrt{1-\frac{\left(da_4/dt_4\right)^2}{c^2}}\right..
\end{equation}

Quantity $a_4(t_4)$ determines in coordinate system $K_4$ distance from any typical observer to the center of $3\mathbb R$-world at the moment $t_4$.
It is the radius of curvature of $3\mathbb R$-world in the coordinate system $K_4$.
The function $a_4(t_4)$ describes the changes radius of curvature of $3\mathbb R$-world in coordinate system $K_4$.
In addition to the geometric meaning, it has also dynamic.
Quantity $da_4/dt_4$ determines the velocity of typical observers in a system $K_4$.
The function $a_4(t_4)$ determines the dynamics of $3\mathbb R$-world in this system.
For example, the distance between typical observers in a system $K_4$ defined by:
\begin{equation}
	\label{22}
	R(t_4)=\left(\frac{R_0}{a_m}\right)a_4(t_4),
\end{equation}
where $R_0=R_4(0)$.

Using \eqref{4}, formula \eqref{555} is written as:
\begin{equation}
\label{13}
	\frac{da_4}{dt_3}=c\sqrt{\left(\frac {a_4}{a_m}\right)^2-1}.
\end{equation}
	
From \eqref{555} we find:
\begin{eqnarray}
\label{14}
\left(\frac{da_4}{dt_3}\right)^2=-c^2+\frac13\Lambda c^2a_4^2,\\
\label{15}
\frac{d^2a_4}{dt_3^2}=\frac13\Lambda c^2a_4,
\end{eqnarray} 
where the constant $\Lambda$ associated with the size $a_m$ by the formula:
\begin{equation}
	\label{16}
	\Lambda=3/a_m^2.
\end{equation}

The system of equations \eqref{14}, \eqref{15} is not correct, because the quantity $t_3$, appearing in these equations are defined in $K_3$, and the quantity $da_4$ and $a_4$ have a specific meaning in the $K_4$ .

Usually, referring to the invariance of the radius of curvature at transition from one coordinate system to another (see, eg, \cite [\S111]{3}), believe that the quantity $a_4(t_4)$ is also the radius of curvature of $3\mathbb R$-world in a system $K_3$ at time $t_3$, associated with $t_4$ by the formula \eqref{19}.
Taking into account the \eqref{6} and \eqref{19}, it is believed that the radius of curvature in a system $K_3$ described by the function
\begin{equation}
	\label{23a}
	a(t_3)=a_4(t_4(t_3))=a_m\cosh\frac{t_3}{t_m}.
\end{equation}
This function is a solution of equations
\begin{equation}
	\label{24a}
	\left(\frac{da}{dt_3}\right)^2=-c^2+\frac13\Lambda c^2a^2,
\end{equation}
\begin{equation}
	\label{25a}
	\frac{d^2a}{dt_3^2}=\frac13\Lambda c^2a^2,
\end{equation}
with initial conditions:
\begin{equation}
	\label{26a}
	a(0)=a_m,\;\;\frac{da}{dt_3}(0)=0.
\end{equation}

Equations \eqref {24a}, \eqref{25a} can be derived from Einstein's equations with $\Lambda$-term, if in it formally set $G=0$, see item \ref{einst_lambda}.
Besides it is necessary to consider that the $3\mathbb R$-world is closed $(k=+1)$.

Equations \eqref{24a}, \eqref{25a} can also be obtained from the equations \eqref{14}, \eqref{15}, if in it remove the badge ``4''.
In the standard derivation of the equations \eqref{24a}, \eqref{25a}, from Einstein's equations with $\Lambda$-term, the idea of an additional fourth spatial dimension is not used.
In this case, there is no need to use badge ``4''.

Taking into account the ideas of the fourth spatial dimension, the dynamics of the $3\mathbb R$-world reasonable for describe by function defining the distance from the center of $3\mathbb R$-world to the typical observer.
In the system $K_4$~---~is $a_4(t_4)$.
In a one-dimensional coordinate system $K_{34}$ it is the distance from the typical observer to the center of this world.
It is described by a function $a_{34}(t_3)$.

To find the function $a_{34}(t_3)$ consider the following problem.
We define the relativistic uniformly accelerated motion, ie rectilinear motion of a particle, which remains with constant acceleration $w$ in their proper (at any given time) reference system related to the particle.

Give a solution to this problem is taken by us from \cite[\S7]{19}.
In the reference frame in which the particle velocity $v=0$ component of 4-acceleration $w^i=d^2x^i/ds^2$ are as follows:
\begin{equation}
	\label{27a}
	w^i=(0,w/c^2,0,0),
\end{equation}
where $w$ --- ordinary three-dimensional acceleration, directed along the axis $x$.
Relativistically invariant condition of uniformly accelerated can be represented as a constant 4-scalar which coincides with $w^2$ in the proper frame of reference:
\begin{equation}
	\label{28a}
	w_iw^i=const=-w^2/c^4.
\end{equation}

In ``fixed'' reference system against which we consider the motion, the disclosure of the expression $w_iw^i$ leads to the equation:
\begin{equation}
	\label{29a}
	\frac d{dt}\frac v{\sqrt{1-\frac{v^2}{c^2}}}=w.
\end{equation}
Integrating this equation with initial conditions $v=0$ for $t=0$ we obtain:
\begin{equation}
	\label{30a}
	v(t)=\frac{wt}{\sqrt{1+\frac{w^2t^2}{c^2}}}.
\end{equation}

Integrating this equation once again and setting $x=x_0=c^2/w$ at $t=0$ we obtain:
\begin{equation}
	\label{31a}
	x=\frac{c^2}w\sqrt{1+\frac{w^2t^2}{c^2}}.
\end{equation}
At $wt\ll c$, these formulas go over into the classical expression $v=wt,\;x=x_0+wt^2/2$.
At $wt\to\infty$ velocity tends to a constant value $c$.

Proper time of uniformly accelerated moving particle
\begin{equation}
	\label{32a}
	t'=\int_0^t\sqrt{1-\frac{v^2}{c^2}}dt=\frac cw \arcsinh \frac {wt}c.
\end{equation}

When $t\to\infty$ it is growing at a much slower than $t$
\begin{equation}
	\label{33a}
	t'\sim\frac cw \ln \frac{2wt}c.
\end{equation}

Expediency of aligning the detailed solution of the relativistic uniformly accelerated motion of a particle is as follows.
Comparison of solutions \eqref{4}, describing the motion typical observer in an inertial system $K_4$ and solutions \eqref{31a}, describing uniformly accelerated motion of a particle in an inertial reference system, shows that they coincide.
This means that the motion of typical observer in a coordinate system $K_4$ can be regarded as uniformly accelerated motion with constant (at any given moment in proper reference system $K_{34}$) acceleration:
\begin{equation}
	\label{34a}
	w=c^2/a_m.
\end{equation}

The value of this acceleration is determined by the initial state of $3\mathbb R$-world.
Acceleration the greater than the minimum size of $3\mathbb R$-world the smaller.
The coordinate system $K_{34}$ is a non-inertial.
The function $a_4(t_4)$ describes the motion of typical observer in an inertial system $K_4$.
In describing the motion of typical observer in a system $K_{4}$ is not necessary to describe the metric properties of the coordinate system $K_{34}$.
Therefore, we present only the relation \eqref{19}, defining the relationship between indication of the clock of a typical observer $t_3$ and the indication of the clock $t_4$ of inertial reference system $K_4$, past which a typical observer flyed.

Function $a_{34}(t_3)$, which describes the motion of the center $3\mathbb R$-world relative to the typical observer can be written in coordinate system $K_{34}$, assuming it immovable.
When this consideration in a coordinate system $K_{34}$, there exists a gravitational field.
Marks of the coordinate system $K_4$ fly past the typical observer with velocity $da_4/dt_4$.
These marks in the location of the typical observer has constant acceleration $w$.

Using symmetry considerations we assume that the laws of motion: a) a typical observer relative to the center of $3\mathbb R$-world in the inertial system $K_4$, b) the center of $3\mathbb R$-world relative to a typical observer in the stationary system $K_ {34}$ are identical.
In view of this the function $a_{34}(t_3)$ defined by the formula:
\begin{equation}
	\label{35a}
	a_{34}(t_3)=\sqrt{a_m^2+c^2t_3^2}.
\end{equation}

For each typical observer of system $K_3$ corresponds a linear system of coordinates $K_{34}$.
All typical observers are equivalent.
All linear coordinates $K_{34}$ are equivalent in describing the motion of the center $3\mathbb R$-world.
The functions $a_{34}(t_3)$, which describe in these systems, the motion of the center of the $3\mathbb R$-world, are identical.

Denote the the function $a_{34}(t_3)$ as $a_3(t_3)$.
This function is a solution of:
\begin{equation}
	\label{25}
	\left(\frac{da_3}{dt_3}\right)^2=c^2-2\Delta(a_3),
\end{equation}
\begin{equation}
	\label{26}
	\frac{d^2a_3}{dt_3^2}=-\frac{d\Delta(a_3)}{da_3},
\end{equation}
where
\begin{equation}
	\label{28}
	\Delta(a_3)=\frac{c^2a_m^2}{2a_3^2}.
\end{equation}

Equations \eqref{25}, \eqref{26} describe the dynamics of $3\mathbb R$-world, when the effect of the gravitational field is not essential.
For this case, these equations are generalized cosmological equations of Friedman (see \ref{rel_model}, equations \eqref{109}, \eqref{110}).
Moreover, the function $a_{34}(t_3)$, not only describes the motion of the center of $3\mathbb R$-world of any of the typical observer, but is also a scale factor of this world in concomitant system of coordinates.

In this approximation, the dynamics of $3\mathbb R$-world is defined by only one parameter --- the minimum size of the world $a_m$.
The parameters $a_m$ and $a_m/c$ are the characteristic spatial and temporal scales, respectively.

Repulsive force is centrifugal, geometric in its nature.
Because the function $a_3(t_3)$ is the length scale in the comoving coordinates system $K_3$, then the distance between any typical observers of system $R_3(t_3)$ in this system are determined by the formula:
\begin{equation}
	\label{x29}
	R_3(t_3)=R_3(0)a_3(t_3)/a_m.
\end{equation}
Obey the Hubble law:
\begin{equation}
	\label{x30}
	\frac{dR_3}{dt_3}=H(t_3)R_3,
\end{equation}
where the Hubble parameter
\begin{equation}
	\label{x31}
	H(t_3)=\frac 1{a_3(t_3)}\cdot\frac{da_3}{dt_3}=\frac{c^2t_3}{a_m^2+c^2t_3^2}.
\end{equation}

\begin{figure*}[h]
\includegraphics [height=6cm]{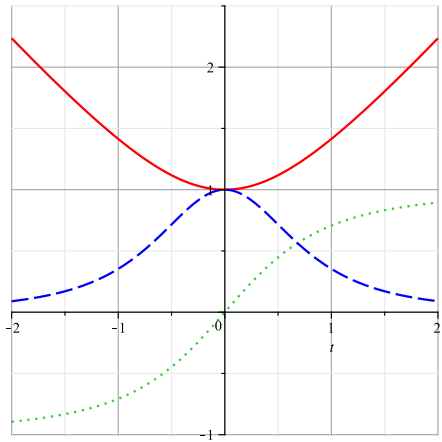} 
\caption{Graphs of the functions describing the change: the scale factor, the rate of its change, as well as cosmological acceleration in the $3\mathbb R$-world. ($a_3(t_3)$ --- the solid line, $da_3/dt_3$ --- a line of dots and $d^2a_3/dt_3^2$ --- a line of dashes).
The unit of distance, time and speed: $a_m$, $t_m$ and $c$, respectively.}
\label{pic1}
\end{figure*}

Graphs of functions $a_3(t_3)$, $da_3/dt_3$ and $d^2a_3/dt_3^2$ are shown in Fig. \ref{pic1}.

In the next section are explained the method of introducing the centrifugal forces of repulsion in the cosmological equations of general relativity.
It is similar to Einstein's method associated to the introduction into these equations $\Lambda$-term.
We briefly describe its essence.

\section{Generalized Einstein equations}
\label{obob_urav_einst}
\subsection{Einstein equations}
The basis of cosmology is general relativity.
According to this theory of four-dimensional space-time in the presence of matter is non-Euclidean.
Metric properties of space-time described by the metric:
\begin{equation}
	\label{p1_1}
	ds^2=g_{ik}dx^idx^k.
\end{equation}
Metric coefficients $g_{ik}$ are functions of the four space-time coordinates ${x_i=(x_0,x_1,x_2,x_3)}$.
They are bijectively related to the distribution of matter and the nature of particle motion of its constituents.
Quantity that determines the properties of matter, is the stress-energy tensor $T_{ik}$.
The relationship between the components of the metric tensor $g_{ik}$ and the stress-energy tensor $T_{ik}$ is determined by the Einstein equations:
\begin{equation}
	\label{p1_2}
	R_i^k-\frac12 \delta_i^kR=\frac{8\pi G}{c^4}T_i^k,
\end{equation}
where $R_i^k$ --- the Ricci tensor, $R$ --- the Ricci scalar, $\delta_i^k$ --- Kronecker delta, see, eg, \cite{1,2,19}.

In cosmology, the space environment is usually described as a continuous ideal solid medium, writing down the stress-energy tensor of the form:
\begin{equation}
	\label{p1_3}
	T_i^k=(\varepsilon+P)u_iu^k-P\delta_i^k,
\end{equation}
where $u_i$ --- the four-velocity of macroscopic motion of the medium, $\varepsilon$ --- the density of the energy and $P$ --- the pressure of the space environment.

\subsection{The geometry of a homogeneous isotropic universe}
In describing the geometry of a homogeneous, isotropic unsteady three-dimensional universe is a space is viewed as a homogeneous and isotropic three-dimensional hypersurface in four-dimensional Euclidean space.

The equation describing the unsteady homogeneous and isotropic three-dimensional hypersurface in the four Cartesian coordinates $(x_1,x_2,x_3,x_4)$ has the form:
\begin{equation}
	\label{p1_4}
	x_1^2+x_2^2+x_3^2+x_4^2=k\cdot a^2(t).
\end{equation}
The signature $k$ takes three values: ${k=+1,-1,0}$.
For $k=1$ realized the case of a space of constant positive curvature.
Value of $k=-1$ corresponds to the space of negative curvature.
Flat space of zero curvature occurs at $k=0$.
Point ${O=(0,0,0,0)}$ is the center of the universe, and $\sqrt{k}\cdot a(t)$~---~its radius.
In a nonstationary universe radius of curvature $a$ varies over time.
We write the metric of spaces with $k=+1,-1,0$ separately.

To describe the geometry of the universe using a three-dimensional curvilinear coordinate system --- comoving systems.
We call it as a coordinate system of typical observers.
Time coordinate is chosen so that in the comoving coordinate system, for any of the typical observer, the interval between two infinitesimally close events occurring at the point where it is determined by the formula:
\begin{equation}
	\label{p1_5}
	ds^2=c^2dt^2.
\end{equation}
Because of the equality of all typical observers introduced so time will be the same for all the observers, and why it is called world time.

For $k=1$ the space of a homogeneous and isotropic universe is three-dimensional hypersphere.
Using four-dimensional spherical coordinates $a,\chi,\theta,\varphi$ (see eg \cite{1,19}), the interval between two infinitesimally close events in the comoving coordinate system is written as:
\begin{equation}
	\label{p1_6}
	ds^2=c^2dt^2-a^2(t)\left(d\chi^2+\sin^2\chi\left[\sin^2\theta(d\varphi)^2+(d\theta)^2\right]\right).
\end{equation}

Formulas describing the geometry of homogeneous spaces of negative curvature, are obtained from describing a spherical universe, if in them are formally replaced $a\to ia $, $\chi\to i\chi$.
At $k=-1$ equation \eqref{p1_4} describes a three-dimensional pseudosphere.
The interval between two infinitesimally close events in comoving coordinates for pseudospherical Universe, is written as:
\begin{equation}
	\label{p1_7}
	ds^2=c^2dt^2-a^2(t)\left(d\chi^2+\sinh^2\chi\left[\sin^2\theta(d\varphi)^2+(d\theta)^2\right]\right).
\end{equation}

Limiting case is when the radius of curvature of three-dimensional space is infinite.
In this case the space of the universe is flat (Euclidean).
In the evolution of a flat universe "frozen"in its Cartesian coordinate system undergoes a homogeneous deformation.
Formally flat space can be described as pseudospherical, writing $ds^2$ as follows:
\begin{equation}
	\label{p1_8}
	ds^2=c^2dt^2-a^2(t)\left(d\chi^2+\chi^2\left[\sin^2\theta(d\varphi)^2+(d\theta)^2\right]\right).
\end{equation}
Detail about the formulas \eqref{p1_6} --- \eqref {p1_8} see, eg, in \cite{1,19}.

\subsection{Cosmological equations of Friedman}
\label{st_ur_fridman}
The metric of a homogeneous, isotropic space contains only one scalar parameter ~ --- radius of curvature $a$.
It defines the curvature of space.
Einstein's equations for a homogeneous, isotropic universe can be converted to the cosmological equation of Friedman defining the relationship between radius of curvature $a$ and the quantities describing the thermodynamic properties of the space matter.

When receiving of the Friedman's equations used by the comoving coordinate system relative to which the medium is at rest, and therefore the components of the four-velocity $u^i=(1,0,0,0)$.
Different from zero are the following components of $T_i^k$:
\begin{equation}
	\label{p1_9}
	T_0^0=\varepsilon,\;\;
	T_1^1=T_2^2=T_3^3=-P.
\end{equation}

Using the expression for the interval between two infinitesimally close events \eqref{p1_6}, \eqref{p1_7}, \eqref{p1_8}, of the Einstein equation \eqref{p1_2} can be transformed to:
\begin{equation}
	\label{p1_10}
	3\left[\left(\frac{\dot a}a\right)^2+\frac{kc^2}{a^2}\right]=\frac{8\pi G}{c^2}\varepsilon,
\end{equation}
\begin{equation}
	\label{p1_11}
	2\frac{\ddot a}a+\left(\frac{\dot a}a\right)^2+\frac{kc^2}{a^2}=-\frac{8\pi G}{c^2}P.
\end{equation}
Hereinafter the dot means a derivative on time $t$.
Equations \eqref{p1_10}, \eqref{p1_11} are called the cosmological equations of Friedman's.
Details for obtaining these equations from Einstein's equations, see, eg, \cite{1}.
Equations \eqref{p1_10}, \eqref{p1_11} can be transformed to the form:
\begin{equation}
	\label{p1_12}
	\frac{d\varepsilon}{da}+3(\varepsilon+P)\frac1a=0,
\end{equation}
\begin{equation}
	\label{p1_13}
	\ddot a=-\frac43\pi G\frac a{c^2}(\varepsilon+3P).
\end{equation}
In a typical space environment $P>0$ and according to the formula \eqref{p1_13}, the effect of the pressure (thermal energy) is not accelerating, but slowing the rate of expansion of the universe.

The view that the thermal energy can only slow down the expansion of a homogeneous universe, is common (see, eg, \cite[Ch.1.]{1}).
The idea about influence of thermal energy in a direction of increasing the rate of expansion of a homogeneous space environment is perceived negatively.
It is usually assumed that, since in a homogeneous medium there is no pressure gradient, it follows that there is no repulsive forces of pressure.
According to the Friedmann equation \eqref{p1_12}, \eqref{p1_13} the thermal energy of a homogeneous, isotropic medium can not change the sign of the cosmological gravitational acceleration, and as can be seen from \eqref{p1_13}, it can only increase the gravity.
Note that this conclusion is derived from the Friedmann equations do not take into account the effect on the space environment cosmological repulsion forces.
In section~\ref{dinamika_onor_rel_mirov} of this work through a simplified evident model shows that in a homogeneous isotropic relativistic universe essential role is played by cosmological centrifugal repulsive force.
They act in the external for the universe fourth spatial dimension and stretch it.
In the comoving coordinate system, these forces are associated with a change in thermal energy of the space environment.
Their effect leads to increase an rate of expansion of of the space environment.
As can be seen from equations of \eqref{p1_13}, these forces within the Einstein equations \eqref{p1_2} can not be described.
To account for the centrifugal cosmological repulsion forces in the equations of GR it is necessary to add terms that describe these forces.

\subsection{Einstein equation with $\Lambda$-term}
\label{einst_lambda}
Variant of the equations GR, which contains a repulsive force was proposed by Einstein \cite{10}.
It is associated with the introduction into gravitational field equations $\Lambda$-term.
In view $\Lambda$-term of the Einstein equation are as follows:
\begin{equation}
	\label{p1_14}
	R_i^k-\frac12 \delta_i^kR=\frac{8\pi G}{c^4}T_i^k+\delta_i^k\Lambda,
\end{equation}
where $\Lambda$ --- the so-called cosmological constant.
In theory, this constant is universal.
Consider that its value can be found by comparing the theoretical predictions and observations.
It is believed that the $\Lambda\approx 10^{-56}\mbox{cm}^{-2}$ \cite{1,8,9}.

Accounting Einstein repulsive forces which are described by the $\Lambda$-term in Einstein's equations, leads to the right side of Friedman's equations \eqref{p1_10} and \eqref{p1_11} with additional terms, and they take the form:
\begin{equation}
	\label{p1_15}
	3\left[\left(\frac{\dot a}a\right)^2+\frac{kc^2}{a^2}\right]=\frac{8\pi G}{c^2}\varepsilon+c^2\Lambda,
\end{equation}
\begin{equation}
	\label{p1_16}
	2\frac{\ddot a}a+\left(\frac{\dot a}a\right)^2+\frac{kc^2}{a^2}=-\frac{8\pi G}{c^2}P+c^2\Lambda.
\end{equation}

The transition from \eqref{p1_2} to \eqref{p1_14} can be formally associated with the replacement:
\begin{equation}
	\label{p1_17}
	T_i^k\Rightarrow T_{i\;eff}^k=(\varepsilon_{eff}+P_{eff})u_iu^k-P_{eff}\delta_i^k,
\end{equation}
\begin{equation}
	\label{p1_18}
	\varepsilon\Rightarrow\varepsilon_{eff}=\varepsilon+\varepsilon_\Lambda,\;\;
	P\Rightarrow P_{eff}=P+P_\Lambda,
\end{equation}
where
\begin{equation}
	\label{p1_19}
	\varepsilon_\Lambda=\frac{c^4\Lambda}{8\pi G},\;\;P_\Lambda=-\varepsilon_\Lambda.
\end{equation}
For details, see, eg, \cite [Ch.4.]{8}.

Often believe that the quantities $\varepsilon_\Lambda$ and $P_\Lambda$ are additions to the $\varepsilon$ and $P$, associated with the existence, besides the usual space environment, some other components of matter, called a ``dark energy''.

From the equations \eqref{p1_15}, \eqref{p1_16} easily obtain a formula which defines the cosmological acceleration  produced by the Einstein's repulsive forces.
It has the form:
\begin{equation}
	\label{p1_20}
	\ddot a_\Lambda=-\frac43\pi G\frac a{c^2}
	\left(\varepsilon_\Lambda+3P_\Lambda\right)=\frac13\Lambda c^2a,
\end{equation}
see, \cite[Ch.4.]{1}.

For Einstein's repulsive force, the sources of which are the quantities $\varepsilon_\Lambda$ and $P_\Lambda$, it is important that $P_\Lambda=-\varepsilon_\Lambda$, owing to what there are forces of repulsion.
Typically pay attention to it.
At the same time, even if $G=0$, the field of Einstein's repulsive force is different from zero (non-recoverable curvature) and, therefore, are essence independent from the gravitational field.
The quantities $\varepsilon_\Lambda$ and $P_\Lambda$ depend from $G$.
The introduction of these variables to describe Einstein's repulsive force is not necessary.
The complexity of the Einstein equations \eqref{p1_14} in practice is the lack of understanding of the physical nature of the $\Lambda$-term.

\subsection{Generalized Einstein's equation for $3\mathbb R$-world}
\label{obob_ur_ein_3r}
Into the equations of general relativity, except for Einstein's repulsion forces associated with $\Lambda$-term, can be introduced others cosmological forces of repulsion \cite{20}.

To introduce these forces in GR, we consider a modification of Einstein's equations of the form:
\begin{equation}
\label{62}
	R_i^k - \frac 12 \delta_i^k R = \frac{8 \pi G}{c^4} T_i^k - \frac{8 \pi \mathbb C}{c^4} Q_i^k,
\end{equation}
where $\mathbb C$-- is a certain constant.
We call these equations the generalized Einstein equations.

Stress-energy tensor of space matter $T_i^k$, defined by \eqref{p1_3} is the source of the gravitational field.
Tensor $Q_i^k$ defined by formula:
\begin{equation}
\label{63}
	Q_i^k=(\varepsilon_\Delta + P_\Delta)u_iu^k - P_\Delta \delta_i^k.
\end{equation}
We believe that it is a source of cosmological repulsion forces.
We assume that the quantities $\varepsilon_\Delta$ and $P_\Delta$ associated with the properties of the space environment and are such that the identity is carried out:
\begin{equation}
\label{64}
	Q_{i;k}^{k} \equiv 0.
\end{equation}
Using the tensor $Q_i^k$, defined by the formula \eqref{63} and satisfying condition \eqref{64}, allows us to introduce into Einstein's equations the additional terms that do not violate the covariance of these equations, and contained in them conservation laws.

In the case of a homogeneous universe $\varepsilon_\Delta$ and $P_\Delta$ defined by the formulas:
\begin{equation}
\label{65}
	\varepsilon_\Delta = \frac {3 c^2}{4 \pi \mathbb C} \frac{\Delta (a)}{a^2} \; ,\;\; P_\varepsilon = 
	- \frac{c^2}{4 \pi \mathbb C } \left (  \frac{\Delta (a)}{a^2} + \frac 1a \frac{d \Delta (a)}{da} \right ),
\end{equation}
where $\Delta(a)$~-- some function of radius of curvature of the Universe $a$.
Quantities $\varepsilon_\Delta$ and $P_\Delta$, as well as $\varepsilon$ and $P$, are scalar functions.
It is easy to verify that for any choice of $\Delta(a)$ the following identity holds: $Q_{i; k}^k\equiv 0$.
Consequently, the conservation laws $T_{i;k}^k=0$ contained in the standard Einstein equations are also present in the equations \eqref{62}.
Arbitrariness in the choice of function $\Delta(a)$ allows you to associate a repulsive force with certain properties of the space environment.

Einstein equation with $\Lambda$-term are a special case of equations \eqref{62}.
Really, if the function $\Delta(a)$ chosen as
$$\Delta (a) = \Delta\Lambda (a) = - \frac 16 \Lambda c^2 a^2 ~,$$
then equation \eqref{62} are the equations of Einstein with $\Lambda$-term.

Given \eqref{63}, \eqref{65}, equations \eqref{62} written in the form:
\begin{equation}
\label{66}
	R_i^k - \frac 12 R \delta_i^k = \frac{8 \pi G}{c^4} T_i^k - \frac {2}{c^2 a^2}
	\left [ \left (3\Delta - \frac {d}{da} (a\Delta)   \right )u_iu^k + \frac {d}{da} (a\Delta)\delta_i^k\right ].
\end{equation}

From the generalized Einstein equations in the standard way we obtain the cosmological equations Friedman.
They are written in the form:
\begin{equation}
\label{p1_26}
	3\left[\frac{\dot{a}^2}{a^2}+\frac{kc^2}{a^2}\right]=\frac{8\pi G}{c^2}\varepsilon-6\frac{\Delta(a)}{a^2},
\end{equation}
\begin{equation}
\label{p1_27}
	2\frac{\ddot{a}}{a}+\frac{\dot{a}^2}{a^2}+\frac{kc^2}{a^2}=-\frac{8\pi G}{c^2}P-\frac{2}{a}\frac{d\Delta(a)}{da}-\frac{2\Delta(a)}{a^2}.
\end{equation}

These equations are called in \cite{20} generalized equations of Friedman.

To the function $\Delta(a)$ may be added an arbitrary constant.
The choice of this constant is determined by the initial conditions (the total energy of the universe).
Formally, addition of constant to $\Delta$-energy means that in cosmological equations of Friedman parameter $k$ can not only adopt the values $(-1,~0,+1)$, but also others. 

System of equations \eqref{p1_26}, \eqref{p1_27} can be transformed to:
\begin{equation}
\label{p1_28}
	\frac{d\varepsilon}{da}+3\left(\varepsilon+P\right)\frac{1}{a}=0,
\end{equation}
\begin{equation}
\label{p1_29}
	\ddot{a}=-\frac{4}{3}\pi G\frac{a}{c^2}\left(\varepsilon+3P\right)-\frac{d\Delta(a)}{da}.
\end{equation}

Equation \eqref{p1_28} is the first law of thermodynamics for a homogeneous isotropic universe.
Assuming that the expansion of the universe is an adiabatic process (see, eg, \cite{1,2}), the first law of thermodynamics is written as:
\begin{equation}
\label{p1_30}
	dE=d\left(\varepsilon V\right)=-P\,dV.
\end{equation}
In a homogeneous isotropic universe, $V\sim a^3$, so \eqref{p1_28} is a consequence of \eqref{p1_30}.

For close the system of equations \eqref{p1_28}, \eqref{p1_29} is necessary to consider the equation describing the thermodynamic properties of the space environment.
In a relativistic universe ($3\mathbb R$-world) pressure $P$ and energy density $\varepsilon$ are related by:
\begin{equation}
\label{p1_31}
	P=\frac13\varepsilon.
\end{equation}

Given \eqref{p1_31}, from \eqref{p1_28} we find that for any $k$ and $\Delta(a)$ we have the equation:
\begin{equation}
\label{p1_32}
	\frac{d \varepsilon}{\varepsilon} + 4\frac{da}{a}= 0.
\end{equation}
Hence we conclude that in a relativistic universe the energy density $\varepsilon$ and the pressure $P$ associated with its radius of curvature of the relations:
\begin{equation}
\label{p1_33}
	\varepsilon a^4\sim Pa^4=const.
\end{equation}

Equation \eqref{p1_29} describes the dynamics of $3\mathbb R$-world in comoving coordinates.
The first term on the right side of this equation defines the forces of gravity, the second - the repulsive forces.
The source of the repulsive forces is ``$\Delta$-energy'' described by the function $\Delta(a)$.
Condition for the appearance of repulsive forces is the presence in the space environment ``$\Delta$-energy'' and its decrease with increasing scale $a$ of $3\mathbb R$-world.

As shown in section \ref{dinamika_onor_rel_mirov}, source of the repulsive forces in an idealized $3\mathbb R$-world, in which ignores the influence of the gravitational field, is the centrifugal energy of radiation (see formula \eqref{25} -- \eqref{28}) .
We assume that this energy is the source of the repulsive forces in a $3\mathbb R$-world is also taking into account the influence of gravitational forces.
``$\Delta$-energy'' into a $3\mathbb R$-world is determined by the formula:
\begin{equation}
\label{p1_35}
	\Delta(a)=\frac{c^2a_m^2}{2a^2}.
\end{equation}

Stress-energy tensor for radiation is written as:
\begin{equation}
\label{p1_34}
	T_i^k=(\varepsilon+P)u_iu^k-\delta_i^kP=\frac\varepsilon 3(4u_iu^k-\delta_i^k).
\end{equation}

Using the formulas \eqref{63}, \eqref{65}, \eqref{p1_35} and \eqref{p1_34}, the summands describing in generalized Einstein equations \eqref{62} cosmological repulsive force is written as:
\begin{equation}
\label{p1_36}
	\frac{8\pi \mathbb C}{c^4}Q_i^k= \frac3{\varepsilon_m a_m^2}T_i^k.
\end{equation}
Cosmological constant of centrifugal forces of repulsion $\mathbb C$ defined by formula:
\begin{equation}
\label{p1_37}
	\mathbb C= \frac{3c^4}{8\pi\varepsilon_m a_m^2}.
\end{equation}
In this case, $Q_i^k=T_i^k$ and the generalized Einstein equation for $3\mathbb R$-world can be written as:
\begin{equation}
	\label{p1_38}
		R_i^k - \frac{1}{2} \delta_i^k R = \frac{8 \pi}{c^4} (G-\mathbb C) T_i^k.
\end{equation}
We assume that the value of the gravitational constant $G$, as the constant of cosmological  repulsive forces $\mathbb C$, is determined by initial state of the $3\mathbb R$-world.
These constants may be different in magnitude.
$3\mathbb R$-worlds in which $G\geq\mathbb C$, generated by the ``Big Bang'' and rapidly expanding.
$3\mathbb R$-worlds in which $\mathbb C>G$, coming from infinity.
First, they are compressed to a certain minimum size, and then accelerated expands to infinity.

\subsection{Generalized Einstein equation for the Universe}
\label{obob_ur_einst_uni}
Given the multicomponent composition of the space environment that fills the universe, to describe it, use two-component approach.
We assume that the environment consists of two uniformly mixed components: non-relativistic and relativistic.

In the nonrelativistic component includes all the components of the space environment, as observed (``baryonic component'') and have not yet observed (``Dark Matter'') consisting of particles, the rest mass which is much larger than their kinetic energy.
Nonrelativistic component is clusterized and is currently the main on mass/energy part of the space environment.
Effect of pressure of the nonrelativistic component on the dynamics of the universe is not essential.

In the relativistic component includes all the components of the space environment, as observed (relict radiation), and are not observed, the equation of state for which $P=(1/3)\varepsilon$.
This component consists of particles with rest mass is equal to zero or much less than their total energy.
We believe that the relativistic component is not clusterized and its spatial distribution is homogeneous.
Currently, the contribution of the relativistic component in total mass/energy space environment is small (see, eg, \cite{1, 2}).
At the same time in the early universe, this contribution was major and determining the dynamics of the universe.

The ratio of concentrations of particles of nonrelativistic $n_M$ and relativistic $n_{rad}$ components, except for the earliest stages of evolution of the universe, remains constant.
According to observational data, $n_{rad}/n_M\sim 10^9$.
Marks $M$ and $rad$ used, as is customary (see, eg, \cite{8}) to denote the quantities describing the nonrelativistic and relativistic component, respectively.

We assume that the source of the cosmological repulsion forces in the universe, as in $3\mathbb R$-world, is the centrifugal energy of the space environment.
The greatest part of this energy is concentrated in a uniformly distributed in the space relativistic component of the space environment.
Given this, the generalized Einstein equation for the universe is written as:
\begin{equation}
	\label{p1_39}
		R_i^k - \frac{1}{2} R \delta_i^k = \frac{8 \pi G}{c^4} (T_{i\,M}^k + T_{i\,rad}^k) - \frac{8 \pi \mathbb C}{c^4} T_{i\,rad}^k.
\end{equation}
Constant $\mathbb C$ is the cosmological constant of the centrifugal repulsive force in the universe.
The value of this constant is determined by the parameters of the initial state of the universe.
It can be found from the practical use of equations \eqref{p1_39}.

The influence of relativistic component of the space environment on the dynamics of the universe is the determining factor in $RD$ (radiation-dominated) era.
In this age of the universe contribution of the relativistic component to the total energy of the space environment is the major (see, eg, \cite{1}, \cite{9}).
Because of the presence of the last term on the right side of the equation \eqref{p1_39}, the dynamics of the universe in $RD$-era may be fundamentally different from the generally accepted.
For example, singularity may be absent in the solutions describing the dynamics of the universe, as well as can take place accelerated expansion (see details in section \ref{prilogenie}).

The influence of the centrifugal energy on the dynamics of of nonrelativistic space environment may also be significant.
Generalized Friedmann equation for an idealized homogeneous of nonrelativistic universe written in the Appendix.

\subsection{Generalized cosmological Friedmann equation}
\label{obob_ur_fridman}

Using the generalized Einstein equation \eqref{p1_39}, in the standard way we transform them into the cosmological Friedmann equation.
For a two-componental space environment, they can be written as:
\begin{equation}
	\label{p1_40}
		3\left(\frac{\dot{a}^2}{a^2}-\frac{k_0 c^2}{a^2}\right) = 8 \pi G (\rho_M + \rho_{rad}) - 8 \pi \mathbb C \rho_{rad},
\end{equation}
\begin{equation}
	\label{p1_41}
		2 \frac{\ddot{a}}{a} + \left(\frac{\dot{a}^2}{a^2}- \frac{k_0 c^2}{a^2}\right) = - \frac{8}{3} \pi G \rho_{rad} + \frac{8}{3} \pi \mathbb C \rho_{rad}.
\end{equation}

In obtaining these equations have neglected the thermal energy of of nonrelativistic component of the space environment.
Believed that the $P_{rad}=(1/3)\varepsilon_{rad}$, energy density $\varepsilon_M$ and $\varepsilon_{rad}$ associated with $\rho_M$ and $\rho_{rad}$ by the equations: $\varepsilon_M=\rho_M\,c^2$, $\varepsilon_{rad}=\rho_{rad}\,c^2$.
The dimensionless parameter $k_0$ is associated with the initial conditions of the universe and determined its total energy.

Equation \eqref{p1_40}, \eqref{p1_41}, can be transformed to:
\begin{equation}
	\label{p1_42}
		\ddot{a} = - \frac{4}{3} \pi G a (\rho_M + \rho_{rad}) - \frac{4}{3} \pi \mathbb C a \rho_{rad},
\end{equation}
\begin{equation}
	\label{p1_43}
		\frac{d}{da} (\varepsilon_M + \varepsilon_{rad})+ (3\varepsilon_M + 4\varepsilon_{rad}) \frac{1}{a}=0.
\end{equation}
Equation \eqref{p1_43} splits into two equations:
\begin{equation}
	\label{p1_44}
		\frac{d \rho_M}{da} + 3 \rho_M \frac{1}{a}=0,
\end{equation}
\begin{equation}
	\label{p1_45}
		\frac{d \rho_{rad}}{da} + 4 \rho_{rad} \frac{1}{a}=0.
\end{equation}

Integrating these equations, we conclude that the density of of nonrelativistic $\rho_M$ and of the relativistic $\rho_{rad}$ components associated with a characteristic size of the universe $a$ by the relations:
\begin{equation}
	\label{p1_46}
		\rho_M (a)= \rho_{M0}(a_0/a)^3 ~,~~ \rho_{rad}(a)=\rho_{rad 0}(a_0/a)^4.
\end{equation}
Here and below zero icon is used to denote the parameters of the present universe.

Given \eqref{p1_46}, the generalized Friedmann equation \eqref{p1_40}, \eqref{p1_41} can be written as:
\begin{equation}
	\label{p1_47}
		\frac{1}{\bar{a}^2} \left( \frac{d \bar{a}}{d \bar{t}}. \right)^2 = k_0 \frac{\Omega_{curv}}{\bar{a}^2} + \frac{\Omega_M}{\bar{a}^3} + \frac{\Omega_{rad}}{\bar{a}^4} (1-\mathbb C /G),
\end{equation}
\begin{equation}
	\label{p1_48}
		\frac{d^2 \bar{a}}{d \bar{t}^2}= - \frac{\Omega_M}{2 \bar{a}^2} - \frac{\Omega_{rad}}{\bar{a}^3} (1-\mathbb C /G),
\end{equation}
where $\bar{a}= a/a_0$, $\bar{t}=t\cdot H_0$, $H_0$~---~Hubble constant.

When writing equations \eqref{p1_47}, \eqref{p1_48} we using the standard notation \cite{8}:
\begin{equation}
	\label{p1_49}
	\Omega_M=\frac{\rho_{M\,0}}{\rho_c},\;
	\Omega_{rad}=\frac{\rho_{rad\,0}}{\rho_c},\;
	\Omega_{curv}=\frac{c^2}{H_0^2a_0^2}.
\end{equation}

Parameters $\Omega_M$ and $\Omega_{rad}$ defined in terms of $\rho_c$ present density of nonrelativistic and relativistic components of the space environment, respectively.

Dimensionless parameter $\Omega_{curv}$ determined by the square ratio of the characteristic length $cH_0^{-1}$ to the scale $a_0$.
Hubble constant is often written in the form: ${H_0=h\cdot 100\,\mbox{km/s Mpc}}$.
In the estimating calculations generally supposed $h=0.7$ \cite{8}.
The critical density $\rho_c$ defined by:
\begin{equation}
	\label{p1_50}
		\rho_c= 3 H_0^2 / 8 \pi G = 1.88 \cdot 10^{-29}h^2 \mbox{g}/ \mbox{cm}^3.
\end{equation}

Solutions of equations \eqref{p1_47}, \eqref{p1_48} satisfy the initial conditions:
\begin{equation}
	\label{p1_51}
		\bar{a}(\bar{t_0})=1,\;(d \bar{a}/ d \bar{t}) (\bar{t_0})=1.
\end{equation}
We believe that the present universe corresponds to the time $t=t_0$.
	
Parameters of equations \eqref{p1_47}, \eqref{p1_48} and the boundary conditions \eqref{p1_51} are:
\begin{equation}
	\label{p1_52}
		\Omega_M,\;\Omega_{rad},\;\Omega_{curv},\; k_0,\;h, \; \mathbb C/G.
\end{equation}
Given the boundary conditions \eqref{p1_51}, from \eqref{p1_47} we find that these parameters are related by correlation:
\begin{equation}
	\label{p1_53}
		k_0 \Omega_{curv} + \Omega_M + \Omega_{rad}(1-\mathbb C /G)=1.
\end{equation}

In the next section will describe a simple cosmological model of the universe, which is based on the equations \eqref{p1_47} and \eqref{p1_48}.

\section{Kinematic cosmological model of the Universe}
\label{rel_model}
In this paper, we restrict the study of solutions of equations \eqref{p1_47}, \eqref{p1_48}, describing the dynamics of the universe without a singularity.

Suppose that initially the scale of the universe had a minimum value $a_m$.
All component of the space environment is in thermodynamic equilibrium and had a temperature $T_m$.
Energy density of relativistic and of nonrelativistic components of this medium were equal $\varepsilon_m$ and $E_m$, respectively.
The condition of minimum initial size means that when $a=a_m,~\dot{a}(a_m)=0$.
Given this initial state, the generalized Friedmann equation \eqref{p1_47}, \eqref{p1_48} can be written as:
\begin{equation}
	\label{p1_54}
		\dot{a}^2=c^2(1-\beta)- \frac{c^2 a_m^2}{a^2} (1-\beta) - c^2\beta \frac {E_m}{\varepsilon_m} \left(1-\frac{a_m}{a}\right),
\end{equation}
\begin{equation}
	\label{p1_55}
		\ddot{a}=\frac{c^2 a_m^2}{a^3} (1-\beta) - \frac{1}{2} c^2 \beta \frac{E_m}{\varepsilon_m} \frac{a_m}{a^2}.
\end{equation}
Dimensionless parameter $\beta$ defined by:
\begin{equation}
	\label{p1_56}
		\beta=G/ \mathbb C =\frac{8}{3}\pi G \frac{\varepsilon_m a_m^2}{c^4}.
\end{equation}

We believe that the temperature of $T_m$ was high enough, but its value was limited.
On the reasonableness of this assumption see, for example, \cite[$\S4$ Ch.6]{1}.
We assume that in the initial state of the universe the energy density of the relativistic component $\varepsilon_m$ was much greater than the energy density of nonrelativistic component $E_m$.

The energy density $\varepsilon_m$ of relativistic component, assuming that it consists of photons and three types of neutrinos can be written as \cite{8}:
\begin{equation}
	\label{ur_4}
		\varepsilon_m = 1.68 \cdot \sigma T_m^4,
	\end{equation}
where is constant Stefan --- Boltzmann 
\begin{equation}
	\label{ur_5}
		\sigma = \frac{\pi^2 k_B^4}{15c^3 \hbar^3},
	\end{equation}
$k_B$ and $\hbar$ is Boltzmann and Planck constants, respectively.
Taking into account \eqref{ur_4}, the constant $\beta$ can be written as:
\begin{equation}
	\label{ur_8}
		\beta=10^{-62} T_m^4 a_m^2,
	\end{equation}
($T_m$ in Kelvin, $a_m$ in centimeters).

Assume that the limiting value of the rate of expansion of the universe is close to the speed of light.
From equation \eqref{p1_54} we see that, $E_m\ll\varepsilon_m$, if the value of $\beta\ll 1$.
Given \eqref{ur_8}, we conclude that $\beta\ll 1$, if
\begin{equation}
	\label{ur_9}
		a_m\ll 10^{31}/T_m^2.
\end{equation}
For example, if $T_m=3\cdot 10^{12}~K$, then inequality \eqref{ur_9} holds for $a_m\ll 10^6\,\mbox{cm}$.

From the equations \eqref{p1_54}, \eqref{p1_55} we see that for the parameters $\beta\ll 1$ and $E_m\ll\varepsilon_m$, the universe in a characteristic time $t_m=a_m/c$ acquires a velocity expansion close the speed of light.
Approximately, the stage of coming out to expanding with speed of light is described by the equations:
\begin{equation}
	\label{109}
		\dot{a}^2=c^2-\frac{c^2a_m^2}{a^2},		
\end{equation}
\begin{equation}
	\label{110}
		\ddot{a}=\frac{c^2 a_m^2}{a^3}.
\end{equation}

These equations coincide with the equations \eqref{25} --- \eqref{28}, describing the motion of the center $3\mathbb R$-world of any typical observer of this world.
This means that the function $a_{34}(t_3)$, describing this movement, also is a scale factor of $3\mathbb R$-world.

Thus, a simple cosmological model of $3\mathbb R$-world, called the kinematic, described in section \ref{dinamika_onor_rel_mirov} present work, is a good approximation to describe the evolution of the universe, if in the initial state of the universe the temperature $T_m$ was high enough and satisfy condition \eqref{ur_9}

Kinematic model of the universe contains only two parameters --- the minimum Size $a_m$, and the maximum temperature $T_m$.
Temperature setting $T_m$ is needed to describe the thermodynamic parameters of the space environment.
Using kinematic models to explain some observational data on the dynamics of the Universe.

\section{Explanation of observational data}
\label{prilogenie}

\begin{itemize}
\item 
In kinematic model of the change of scale of the universe in comoving coordinates is given by
\begin{equation}
	\label{112}
		a(t)= \sqrt{a_m^2 + c^2 t^2}.
\end{equation}

\item 
According to this model, the Universe very quickly, during characteristic time $t_m=a_m/c$, has acquired the velocity of expansion is close to the speed of light.
Have long had an almost uniform expansion of the universe.
So, apparently, is not by chance a good approximation to describe its dynamics is an S-model (model of a uniformly expanding universe), proposed in \cite{20}.

\item 
Considering that the age of the universe $t_0\gg a_m/c$ and using \eqref{112}, from \eqref{x31} we get:
\begin{equation}
	\label{111}
		t_0 \approx H(t_0)^{-1}=H_0^{-1}.
\end{equation}
Since the Hubble constant $H_0\approx 70\,\mbox{km/s\,Mpc}$, then $t_0\approx 14\cdot 10^9\,\mbox{years}$.
This gives a correct estimate of the age of the universe.
The scale of the present universe $a_0\approx c\,t_0\approx 10^{28}\,\mbox{cm}$.

\item
Using \eqref{112} we find the formula that determines in the open universe a photometric distance to typical observers with redshift $z$:
\begin{equation}
	\label{113}
		\bar{r}(z)= \frac{r(z)}{cH_0^{-1}}=\frac{1}{\sqrt{\Omega_{curv}}} \sinh \int^{z}_{0}{\frac{d\acute{z}}{(1+\acute{z})\sqrt{1-(a_m^2/a_0^2) (1+\acute{z})^2}}},
\end{equation}
where
\begin{equation}
	\label{114}
		\Omega_{curv} = \frac{c^2}{a_0^2 H_0^2} = \frac{a_0^2}{a_0^2 - a_m^2}.
\end{equation}
Details of obtaining the formula \eqref{113} see, for example, in \cite[Ch.4]{8}.
\item 
Assuming that the characteristic expansion time $t_m$ is much smaller than the age of the universe at the time of recombination, we believe that a good approximation for explaining the known observational data give the formula:
\begin{equation}
	\label{115}
		a(t)=ct ~,~~\Omega_{curv}=1 ~,~~ \bar{r}(z)= \sinh \ln(1+z)= \frac {2z+z^2}{2(1+z)}.
\end{equation}

\item 
Taking into account \eqref{115}, the formula determining the dependence of ``apparent magnitude --- redshift'' for objects that have a certain absolute luminosity can be written as:
\begin{equation}
	\label{116}
		(m-M)(z)=5 \lg \left(z+\frac{z^2}{2}\right) + 5 \lg \left( \frac{c H_0^{-1}}{l_0}\right),
\end{equation}
$l_0=10 \mbox{pc}$ (see, eq, \cite{20}).
Figure 2 shows the function $(m-M)(z)$, calculated by formula \eqref{116}, and observational data for supernova type Ia, with, presumably, the same absolute luminosity at the peak of their brilliance.
Can see that function \eqref{116} well explains the observation


\begin{figure}[h]
\includegraphics[width=10cm]{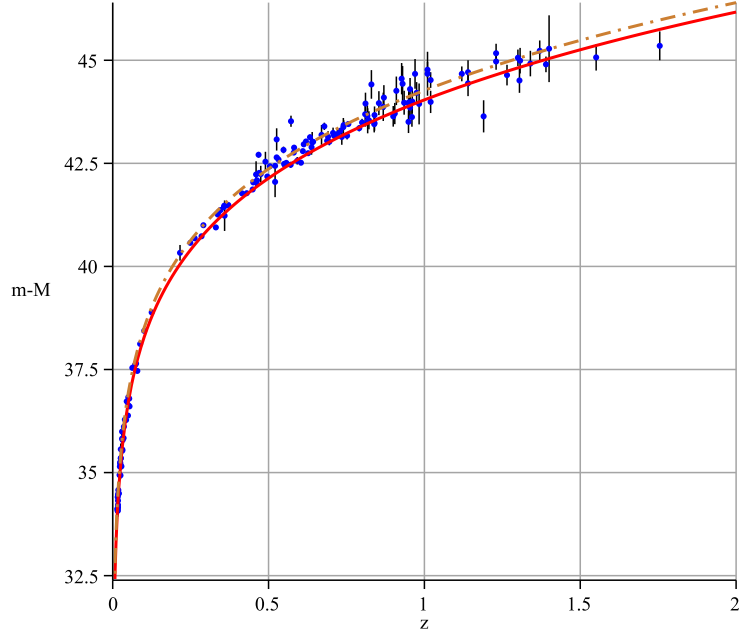}
\caption{
The graph of ``apparent magnitude --- redshift'' $(m-M)(z)$ for objects having a certain absolute luminosity.
Graphics were obtained in the kinematic model (for $H_0=70\,\mbox{km/s\,Mpc}$~---~ solid line; $H_0=63\,\mbox{km/s\,Mpc}$~---~ line of dots and dashes).
The dots are observed values of $(M-m)(z)$ for supernovae of type Ia, vertical intervals~--- error of their determination \cite{5, 6}.
}
\label{pic2}
\end{figure}

\item 
Taking into account \eqref{115}, the formula determining the angular Size (degrees) of the observed object that has a redshift of $z$ and the linear Size $d$, can be written as:
\begin{equation}
	\label{117}
		\Delta \theta = \frac {2d(1+z)^2}{(2z+z^2)cH_0^{-1}} \frac{180}{\pi},
\end{equation}
(
see, eg, \cite[Ch.4]{8}). 
Using \eqref{117}, we find that objects with linear dimensions $d$ and $z\gg 1$ seen at an angle
\begin{equation}
	\label{118}
		\Delta \theta = \frac{2d}{cH_0^{-1}} \frac{180}{\pi}.
\end{equation}

\item 
On a uniform background of cosmic relict radiation observed bright spots with angular Size roughly equal to one degree (see,for example \cite{7,8,9}).
Given \eqref{118}, we conclude that objects with $z\gg 1$ and the observed angle $\Delta\theta=1^\circ$, are of size
\begin{equation}
	\label{119}
		d_1 = c H_0^{-1} \frac{\pi}{360} \approx 40 \mbox{Mpc}.
\end{equation}
This size is, apparently, is the maximum size of inhomogeneities are observed in the universe, see, eg, \cite{1}.
If we assume that inhomogeneities are gravitationally bound systems and are not affected by the cosmological expansion, then $d_1$ defines a linear size of bright spots that visible on a uniform background of relict radiation.

\item 
Typically, the linear dimensions of the observed bright spots on a uniform background of cosmic microwave background radiation associated with the size of cause-related areas at the time of recombination.
We find  the moment of recombination $t_{rec}$ in the kinematic model of the universe is given that the universe is expanding uniformly, and following relation holds:
\begin{equation}
	\label{120}
		t_0/t_{rec}=T_{rec}/T_0.
\end{equation}
The present value of the CMB temperature $T_0\approx 3\,K$, and the temperature at the time of recombination $T_{rec}\approx 3\cdot 10^3\,K$.
Using \eqref{120}, we obtain:
\begin{equation}
	\label{121}
		t_{rec} \approx 10^{-3}t_0 \approx 14 \cdot 10^6 \mbox{y.}.
\end{equation}
The size of cause-related areas at the time of recombination is equal to:
\begin{equation}
	\label{122}
		d_{rec}= 2 c t_{rec} \approx 28 \cdot 10^6 \mbox{ly} \approx 9 \mbox{Mpc}.
\end{equation}

\item 
The size $d_{rec}$ much less than $d_1$.
This means that the observed bright spots against a uniform background of relict radiation appeared considerably later than the time of recombination.
Inhomogeneity having a size of $\approx 40\,\mbox{Mpc}$, must have redshifts $z$ is not greater than $200\div 220$.
It is possible that this disparity between the $d_1$ and $d_{rec}$ indicates the need to refine the model.

\end{itemize}

\section{Conclusion}
\label{zakl}
\begin{enumerate}
\item 
To qualitatively explain the nature of the cosmological repulsion forces, we consider the dynamics of an idealized isotropic homogeneous three-dimensional world filled with radiation ($3\mathbb R$-world.)
It is believed that $3\mathbb R$-world consists of freely-moving photons and it is immersed in a four-dimension flat Euclidean space.
At first, he has a minimum scale $a_m$ and a maximum temperature $T_m$.

\item
Shown that the hypothesis of the fourth large-scale spatial dimension has an important consequence.
Even in the case of free motion of particles, $3\mathbb R$-world is expanding accelerated.
In the framework of continuum mechanics, this extension can be considered as occurring under the action of repulsive forces.
These forces are centrifugal by its nature.
They acting in an external for $3\mathbb R$-world fourth spatial dimension and stretch the world.
Centrifugal forces have no effect on the thermodynamic properties of the space environment.

\item 
Equations are found that describe the dynamics of $3\mathbb R$-world, consisting of freely moving relativistic particles in a comoving coordinate system (equation \eqref{25}---\eqref{28}).
$3\mathbb R$-world is expanding accelerated and it is interpreted as the result of the repulsive forces.
With increasing size of $3\mathbb R$-world, these forces decrease rapidly.
The amplitude of the repulsive force $c^2/a_m$ is greater then the minimum size of $3\mathbb R$-world $a_m$ is smaller.
At the same time, the lower value of size of $3\mathbb R$-world $a_m$, corresponds a more rapid decrease of repulsive force.
When size $a$ much larger then the size $a_m$ then the expansion of $3\mathbb R$-world occurs at a speed close to the speed of light.
It is described by the function
\begin{equation}
	\label{z_1}
		a(t)= \sqrt{a_m^2 + c^2 t^2}.
\end{equation}

\item 
It is shown that the centrifugal forces in homogeneous isotropic infinite gravitating environments can not be described within the standard equations of GR.
To describe these forces requires a generalization of Einstein's equations.

\item 
Put forward the hypothesis that the source of the cosmological repulsion forces in the universe is the centrifugal energy of the space environment.
Much of this energy is concentrated in a uniformly distributed in the space of the relativistic component of the space environment.
Taking this into account, are written down generalized Einstein equations, considering the effect of centrifugal forces on the dynamics of the universe. 

\item 
Written down the generalized cosmological Friedmann equation describing the evolution of a homogeneous isotropic universe with the influence of the centrifugal energy of the space environment.
Investigated solutions of these equations, which describe the universe without singularities and without divergence rate of its expansion.
Considered the case when the asymptotic value of expansion rate  equals the speed of light.
Approximating the dynamics of the universe is described by the equation \eqref{z_1}.

\item 
Approximating the dynamics of the universe is described by the equation \eqref{z_1}
This model assumes that initially the universe had a minimum Size $a_m$ and a maximum temperature $T_m$.
The energy density of relativistic components of the space environment was much greater than the energy density of nonrelativistic components.
The minimum Size of the universe $a_m$ is sufficiently small.
The condition \eqref{ur_9} was satisfied.

\item 
Kinematic cosmological model of the universe contains only two parameters: the minimum Size of the universe $a_m$ and its maximum temperature of $T_m$.
According to this model, the Universe very quickly, in a characteristic time $t_m=a_m/c$, got the expansion velocity is close to the speed of light.
If $a\gg a_m$, then a good approximation to describe its evolution is a model of a uniformly expanding of the universe.

\item 
Using a kinematic model to explain some important observational data about the dynamics of of the universe shows the following.
\begin{itemize}
\item 
The model correctly explains the age of the universe.
\item 
It gives a simple explanation for the observed dependence ``visible stellar magnitude~--- redshift'' for supernovae type Ia
\item 
Model allows us to make certain opinions about the characteristic size and time of occurrence of the observed inhomogeneities in a uniform background of relict radiation.
\end{itemize}

\end{enumerate}

\begin{acknowledgments}
Authors express profound gratitude I.G.~Shukhman for the numerous useful discussions of  problem solved in the present work.
We are grateful to A.G.~Zhilkin and A.~Frenkel for valuable advice and helpful comments.
\end{acknowledgments}

\appendix
\section{Nonrelativistic universe}
\label{pril1}
Consider the following idealized system.
Homogeneous and isotropic space uniformly filled gravitating nonrelativistic monatomic ideal gas.
Using the generalized cosmological Friedmann equation and the equations describing the thermodynamic properties of an ideal gas, study the dynamics of this system.

Standard cosmological Friedmann equations \eqref{p1_10}, \eqref{p1_11} contains terms describing the kinetic energy of radial motion of the space environment $\sim\dot{a}^2$, but not contain analogous in the form of terms describing the energy of thermal motion .
To describe the dynamics of nonrelativistic universe using the generalized Friedmann equation \eqref{p1_26}, \eqref{p1_27}.
We assume that the description of the energy of the radial and the thermal motion of the space environment in these equations should be the same type.
This description can be achieved if the function $\Delta(a)$ take the form:
\begin{equation}
\label{P1}
	\Delta(a)=\frac 1 2 v^2(a) - \frac 1 2 c^2 (k+k_0),
\end{equation}
where $v^2(a)/2$~--- the energy of thermal motion of unit mass of the space matter.
The choice of the constant $k_0$ is determined by initial conditions.

Taking into account \eqref{P1} the generalized cosmological equations A. Friedman can be written as:
\begin{equation}
\label{P2}
	3\left(\frac{\dot{a}^2}{a^2}-\frac{k_0\,c^2}{a^2}\right)=\frac{8\pi G}{c^2}\varepsilon-\frac{3\,v^2}{a^2},
\end{equation}
\begin{equation}
\label{P3}
	2\frac{\ddot{a}}{a}+\left(\frac{\dot{a}^2}{a^2}-\frac{k_0\,c^2}{a^2}\right)=-\frac{8\pi G}{c^2}P-\frac{1}{a}\frac{dv^2}{da}-\frac{v^2}{a^2}.
\end{equation}

Differentiating the equation \eqref{P2} for $t$, we find:
\begin{equation}
\label{P4}
	\ddot{a}=\frac{4\pi G}{3\,c^2}\frac{d}{da}\left(\varepsilon\,a^2\right)-\frac{1}{2}\frac{dv^2}{da}.
\end{equation}

From equation \eqref{P2}, \eqref{P3}, we get:
\begin{equation}
\label{P5}
	\ddot{a}=-\frac{4}{3}\frac{\pi G\,a}{c^2}(\varepsilon+3\,P)-\frac{1}{2}\frac{dv^2}{da}.
\end{equation}
It is seen that the thermal energy is not only one of the sources of the gravitational field that takes into account due to the dependence $\varepsilon$ and $P$ from $v^2$, but at the same time and cause of the repulsive forces.

Equating \eqref{P4}, \eqref{P5}, we conclude that for any kind of functions $v^2(a)$ and parameter $k$ we have an equation describing the conservation of energy of the space environment in an adiabatic process:
\begin{equation}
\label{P6}
	\frac{d\varepsilon}{da}+3(\varepsilon+P)\frac{1}{a}=0.
\end{equation}

For non-relativistic monatomic gas dependence $v^2(a)$ is quite defined and is a consequence of the equation \eqref{P6}.
For non-relativistic monatomic ideal gas we have the formulas:
\begin{equation}
\label{P7}
	\varepsilon=n\,m_0c^2+\frac{1}{2}n\,m_0v^2=\varepsilon_0+\varepsilon_k,
\end{equation}
\begin{equation}
\label{P8}
	\varepsilon_k=\frac{1}{2}n\,m_0v^2=\frac{3}{2}n\,k_BT,\;P=n\,k_BT=\frac{2}{3}\varepsilon_k,
\end{equation}
where $n$~--~concentration of particles, $\varepsilon_k$~--~thermal energy per unit volume, $T$~--~temperature, $P$~--~pressure of an ideal gas, $k_B$~--~Boltzmann constant.

It is easy to show that the equation \eqref{P6} splits into two:
\begin{equation}
\label{P9}
	\frac{dn}{da}+3\,n\frac{1}{a}=0,
\end{equation}
\begin{equation}
\label{P10}
	\frac{d\varepsilon_k}{da}+\frac{5\,\varepsilon_k}{a}=0.
\end{equation}

Equation \eqref{P9} is the law of conservation of particle number.
From this it follows:
\begin{equation}
\label{P11}
	n\,a^3=const=n_0a_0^3.
\end{equation}

Equation \eqref{P10} describes the law of conservation of thermal energy in an adiabatic process.
\begin{equation}
\label{P12}
	\varepsilon_ka^5=const.
\end{equation}

Since $\varepsilon_k\sim n\,v^2$, in view of \eqref{P11}, from \eqref{P12}, we obtain:
\begin{equation}
\label{P13}
	v^2a^2=const=L_0^2=v_0^2a_0^2.
\end{equation}

Given the smallness of the parameter $v^2/c^2$, the equation \eqref{P5} is simplified and takes the form:
\begin{equation}
\label{P14}
	\ddot{a}=-\frac{4}{3}\pi Ga\rho-\frac{1}{2}\frac{dv^2}{da}.
\end{equation}

Using \eqref{P11} and \eqref{P13}, equation \eqref{P14} is written as:
\begin{equation}
\label{P15}
	\ddot{a}=-\frac{GM_0}{a^2}+\frac{L_0^2}{a^3}=-\frac{dU_{eff}}{da},
\end{equation}
where
\begin{equation}
\label{P16}
	U_{eff}=-\frac{GM_0}{a}+\frac{L_0^2}{2a^2},
\end{equation}
\begin{equation}
\label{P17}
	M_0=M(a)=\frac{4}{3}\pi\rho a^3=\frac{4}{3}\pi\rho_0a_0^3.
\end{equation}

\begin{figure*}[h]
\includegraphics[width=10cm]{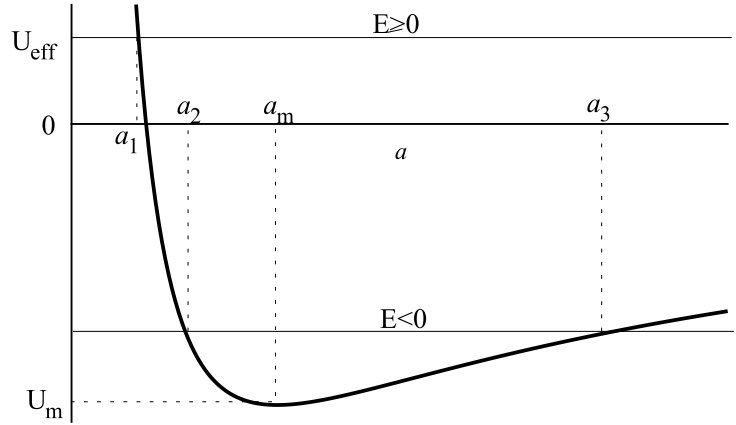}
\caption{Graph of the function $U_{eff}(a)$, defined by \eqref{P16}.}
\label{pic3}
\end{figure*}

Graph of the function $U_{eff}(a)$ is shown in Fig.\ref{pic3}.
The quantities $M_0$ and $L_0^2$ are the parameters of the idealized non-relativistic universe.

First integral of the equation \eqref{P15} is the energy:
\begin{equation}
\label{P18}
	E=\frac{\dot{a}^2}{2}-\frac{GM_0}{a}+\frac{L_0^2}{2a^2}.
\end{equation}
The energy $E$ and parameter $k_0$ are related by $E=k_0c^2/2$.

Equation \eqref{P18} is the law of conservation of energy nonrelativistic space environment.
In contrast to the corresponding conservation law ``standard'' Friedmann equations, in the equation \eqref{P18} is taken into account that the changing of kinetic energy of radial motion of the space environment is connected not only to the changing of potential energy, but also  with changing of energy of thermal motion.
Energy of thermal motion of the medium $v^2/2$, in our problem, is both the centrifugal energy.
Satisfies the relation:
\begin{equation}
\label{P20}
	v^2/2=L_0^2/2a^2.
\end{equation}

Equation \eqref{P15} differs from the ``standard'' equation Friedman, for a nonrelativistic medium, the presence of the right side of the term $L_0^2/a^3$, describing the action on the space environment of centrifugal forces.
These forces exist due to the presence of thermal energy of the space environment.

Equation \eqref{P15} is similar to the equation describing the radial motion of unit mass, with angular momentum $L=L_0$, in the gravitational field of a point mass $M_0$.
Solutions describing the dynamics of the universe will be radically different depending on whether $E<0$ or $E\geq 0$.

\begin{figure*}[h]
\includegraphics[width=10cm]{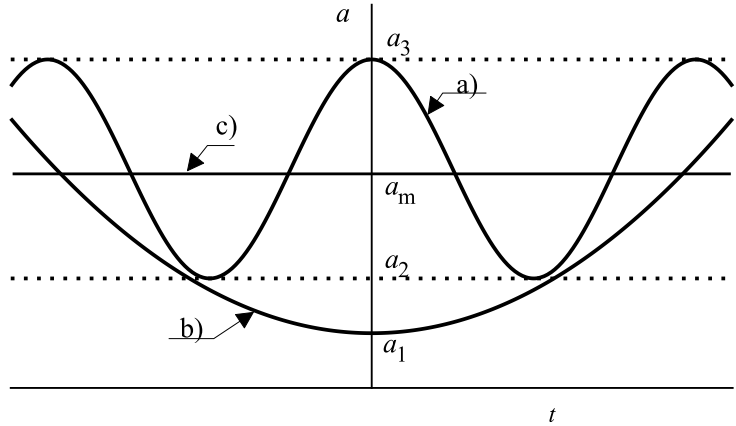}
\caption{Possible types of solutions of \eqref{P15}, describing:   a)---oscillating universe $(E<0)$; b)---open universe $(E\geq0)$; c)---stationary universe $(E=U_m)$.}
\label{pic5}
\end{figure*}

If $E<0$ then nonrelativistic idealized universe is closed and oscillating.
It is described by solutions of the type a) (see Fig. \ref{pic5}.), has finite volume and mass.

If $E\ge 0$, the universe is open.
Describe the evolution of such a universe is a infinite solutions of type b) (see Fig.\ref{pic5}.).

When performing ``initial'' conditions:
\begin{equation}
\label{P21}
	a_0=a_s=\frac{L_0^2}{GM_0},\;\dot{a}(0)=0
\end{equation}
realized by a stationary solution $a=a_s$ (type c), see Fig.\ref{pic5}.).
The universe in this case is closed, has finite volume $V=2\pi^2a_s^3$.
Stationary state nonrelativistic universe is gravitationally stable.

In non-relativistic universe with $L_0\ne 0$ there is no singular state.
In this universe ``Big Bang'' can occur only if $L_0=0$.

Conditions $E<0$ and $E\ge 0$ may be written in a form that is commonly used in cosmology, namely: $E<0\Rightarrow\rho_0>\rho_c$, $E\ge 0\Rightarrow\rho_0\le\rho_c$, respectively.
In the considered here model, the role of the critical density is played by quantity
\begin{equation}
\label{P22}
	\rho_c=\frac{3}{8\pi G}\left(H_0^2+\frac{L_0^2}{a_0^4}\right).
\end{equation}
This formula shows that the cosmological centrifugal forces significantly affect  the value of the critical density.

\end{document}